\documentclass{PoS}

\title{Dying radio galaxies in the LOFAR Lockman Hole}

\ShortTitle{Dying radio galaxies in the LOFAR Lockman Hole}

\author{\speaker{Marisa Brienza}\\
        ASTRON, the Netherlands Institute for Radio Astronomy, Postbus 2, 7990 AA, Dwingeloo, NL\\
        Kapteyn Astronomical Institute, University of Groningen, Landleven 12, 9747 AD Groningen, NL\\
        E-mail: \email{brienza@astron.nl}}
\author{Elizabeth Mahony\\
		ASTRON, Netherlands Institute for Radio Astronomy, Postbus 2, 7990 AA, Dwingeloo, NL\\
		Sydney Institute for Astronomy, School of Physics A28, The University of Sydney, NSW 2006, AU\\
		ARC Centre of Excellence for All-Sky Astrophysics (CAASTRO)\\}
\author{Raffaella Morganti\\
		ASTRON, Netherlands Institute for Radio Astronomy, Postbus 2, 7990 AA, Dwingeloo, NL\\
        Kapteyn Astronomical Institute, University of Groningen, Landleven 12, 9747 AD Groningen, NL\\}
\author{Isabella Prandoni\\
		INAF-ORA Bologna, Via P. Gobetti 101, 40129 Bologna, Italy\\}
\author{Leith Godfrey\\
		ASTRON, Netherlands Institute for Radio Astronomy, Postbus 2, 7990 AA, Dwingeloo, NL\\}

\abstract{After the jets have switched off, radio galaxies undergo a fading phase which is often named the dying phase. The luminosity evolution of the remnant plasma during this period is still poorly constrained because of the paucity of objects detected. Using the new 150-MHz deep LOFAR observations of the well-known extragalactic field the Lockman Hole, we performed a systematic search of dying radio sources aiming to provide good statistics on their detection and properties. To avoid selection biases towards any specific class of dying sources we used both morphological and spectral selection criteria. To do this we combined the LOFAR data with publicly available surveys at other frequencies as well as dedicated deep observations. Our preliminary results, show that the fraction of candidate dying radio sources is < 6-8\% of the entire radio source population and is dominated by steep spectrum sources. By comparing these observational results with statistical modelling of the radio sky population we will be able to constrain the main mechanisms contributing to the dying radio galaxy luminosity evolution. }

\FullConference{EXTRA-RADSUR2015 (*)\\
		20--23 October 2015\\
		Bologna, Italy

                \bigskip
                \hrule
                \bigskip

                \textnormal{(*) This conference has been organized
                  with the support of the Ministry of Foreign Affairs
                  and International Cooperation, Directorate General
                  for the Country Promotion (Bilateral Grant Agreement
                  ZA14GR02 - Mapping the Universe on the Pathway to
                  SKA)}
}

\begin{document}

\section{Introduction}

The active phase of radio-loud active galactic nuclei (AGN) can typically last several tens of Myr (Parma et al. 1999). After the radio jets switch off the source starts to fade away and is usually termed dying or remnant radio galaxy. The fate of the AGN remnant radio plasma and the physical processes driving its evolution have implications for several areas of radio galaxy research. Firstly, the modelling of their radio spectrum provides constraints on the time-scales of activity and quiescence of the radio source (Kardashev 1962; Pacholczyk 1970; Jaffe \& Perola 1973). This can help in understanding the life-cycle of radio galaxies as well as quantifying the role of radio AGN feedback. Moreover, it can give new insights into the formation of radio sources in galaxy clusters like relics, halos and phoenixes (En{\ss}lin et al. 2002). 

The current knowledge about dying radio galaxies is mainly based on detailed analysis of individual sources that have been serendipitously identified (e.g. Jamrozy et al. 2004). A few searches have been conducted (see Sec. 3) but systematic studies of dying radio galaxies as a population have been prevented by the small number of sources detected. The observed fraction stands well below the predictions of radio galaxy evolution models. The reason for this rarity may be both connected to a much more rapid luminosity evolution than expected and/or the difficulty of their identification. 

For a long time there have been claims that deep low-frequency surveys would have enhanced the detection of this class of sources. Indeed, preferential radiative cooling of particles at high frequency makes the remnant plasma more likely to be detectable in the MHz-regime. Thanks to its high sensitivity and complete uv-coverage, the Low-frequency Array (LOFAR, van Haarlem et al. 2013) is currently the best instrument to detect low surface brightness emission at low frequency and with high resolution. In particular, the continuum survey, which is one of the Key Science Project (KSP) of LOFAR (R\"ottgering et al. 2003, van Haarlem et al. 2013) is currently mapping the entire northern sky with a resolution of 5 arcsec and an rms of 0.1 mJy/beam at frequencies 120-180 MHz. 

In this context, we started a systematic search for dying sources in one of the largest and well-characterized extragalactic deep fields, i.e. the Lockman Hole. This work is motivated by the recent observations at 150-MHz performed by Mahony et al. (in prep.) as part of the Continuum Surveys KSP. By having a multitude of radio data (15 GHz, Whittam et al. 2013; 1.4 GHz, Guglielmino et al. 2012, Prandoni et al. in prep.; 600 MHz, Garn et al. 2010; 345 MHz, Prandoni et al. in prep.) this field is best for testing our search strategy that will then be exported to other fields.

\section{The Lockman Hole field at 150 MHz}

The Lockman Hole field (centred at RA=10:47:00 and Dec=58:05:00 J2000) was observed with the LOFAR High-Band Antennas (HBA) for 10 hrs on March 18, 2013. Only the Dutch stations were used for these observations resulting in baselines ranging from $\sim$40 m up to $\sim$120 km. The data were calibrated and imaged using the standard LOFAR tools (see Heald et al. 2010) with the amplitude scale set according to Scaife \& Heald (2012). The observations covered the frequency range 110-182 MHz, but to maximise the sensitivity the full bandwidth was combined to form a single image with a central frequency of 150 MHz.

The resulting image (see Fig. 1, top panel) has a final beam size of 18.6$\times$14.7 arcsec and an rms of 150 $\rm \mu$Jy/beam at the centre of the field. Source extraction was carried out using PyBDSM (Mohan \& Rafferty 2015) which resulted in a catalogue of 5859 sources above a flux limit of 0.75 mJy. Full details on the data reduction and source extraction will be published in the LOFAR catalogue paper on the Lockman Hole field (Mahony et al, in prep.), but further details can also be found in Mahony et al. 2015.

In order to investigate the spectral indices of these radio sources, the 150~MHz LOFAR catalogue was cross matched with deep 1.4~GHz Westerbork observations (Guglielmino et al. 2014). The Westerbork image covers an area of $\sim$7 sq. degrees (compared to the $\sim$30 sq. degree field of the LOFAR observations) at a similar resolution of 11$\times$9 arcsec and reaches an rms noise of 11 $\rm \mu$Jy/beam. Cross matching these two catalogues resulted in a sample of 1366 sources with a median spectral index between 150 MHz and 1.4 GHz of $\alpha$=0.8, typical of low-frequency selected surveys (Intema et al. 2011, Williams et al. 2013). Further cross matches with higher-frequency catalogues will be presented in Mahony et al. (in prep.). 

To expand the spectral study to the entire LOFAR field-of-view, we also re-imaged the LOFAR data with a uv-cut at 4 k$\rm \lambda$ and a final restoring beam of 45 arcsec (see Fig. 1 bottom panel) to match the resolution of the 1.4-GHz NRAO VLA Sky Survey (NVSS, Condon et al. 1998). The rms at the centre of the field is 1.2 mJy/beam over the full-bandwidth. The source extraction was performed with PyBDSM and the catalogue contains 2588 sources above a flux limit of 6 mJy. We then cross matched this low-resolution LOFAR catalogue with the NVSS catalogue. For consistency, and to form a complete sample, we only included sources with $\rm S_{150}$>36.5 mJy in the low-resolution LOFAR catalogue meaning that any LOFAR sources without an NVSS counterpart have steeper spectra than $\rm \alpha$=1.2. Sources that do not have a counterpart in the NVSS catalogue have been assigned an upper limit flux density of 2.5 mJy equal to the flux limit of the NVSS. By making this cut the catalogue is restricted to 743 sources and the spectral index distribution peaks at $\rm \alpha$=0.77, in good agreement with the high resolution catalogue.

\begin{figure}[h!]
\centering
\includegraphics[width=0.8\textwidth]{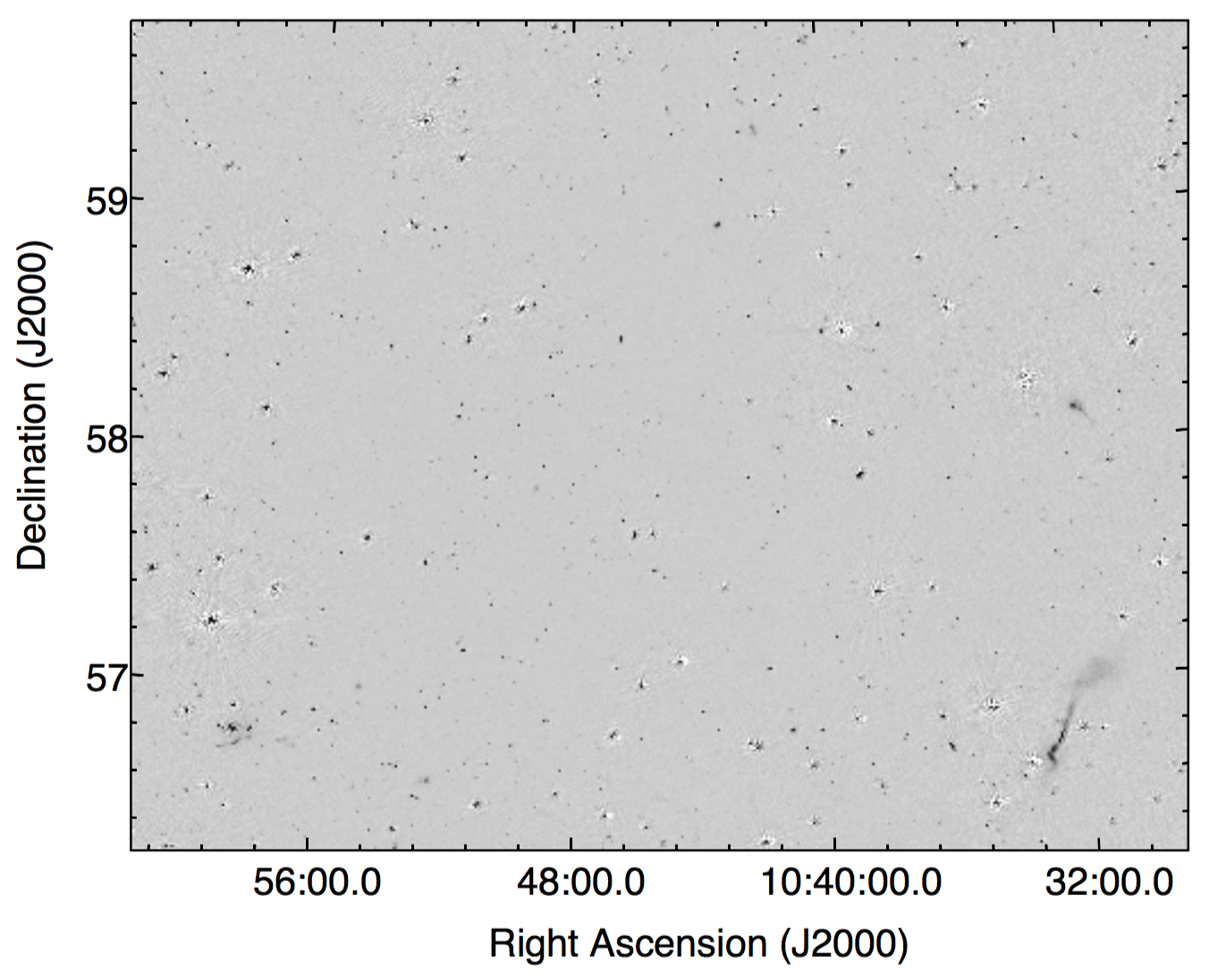}
\includegraphics[width=0.8\textwidth]{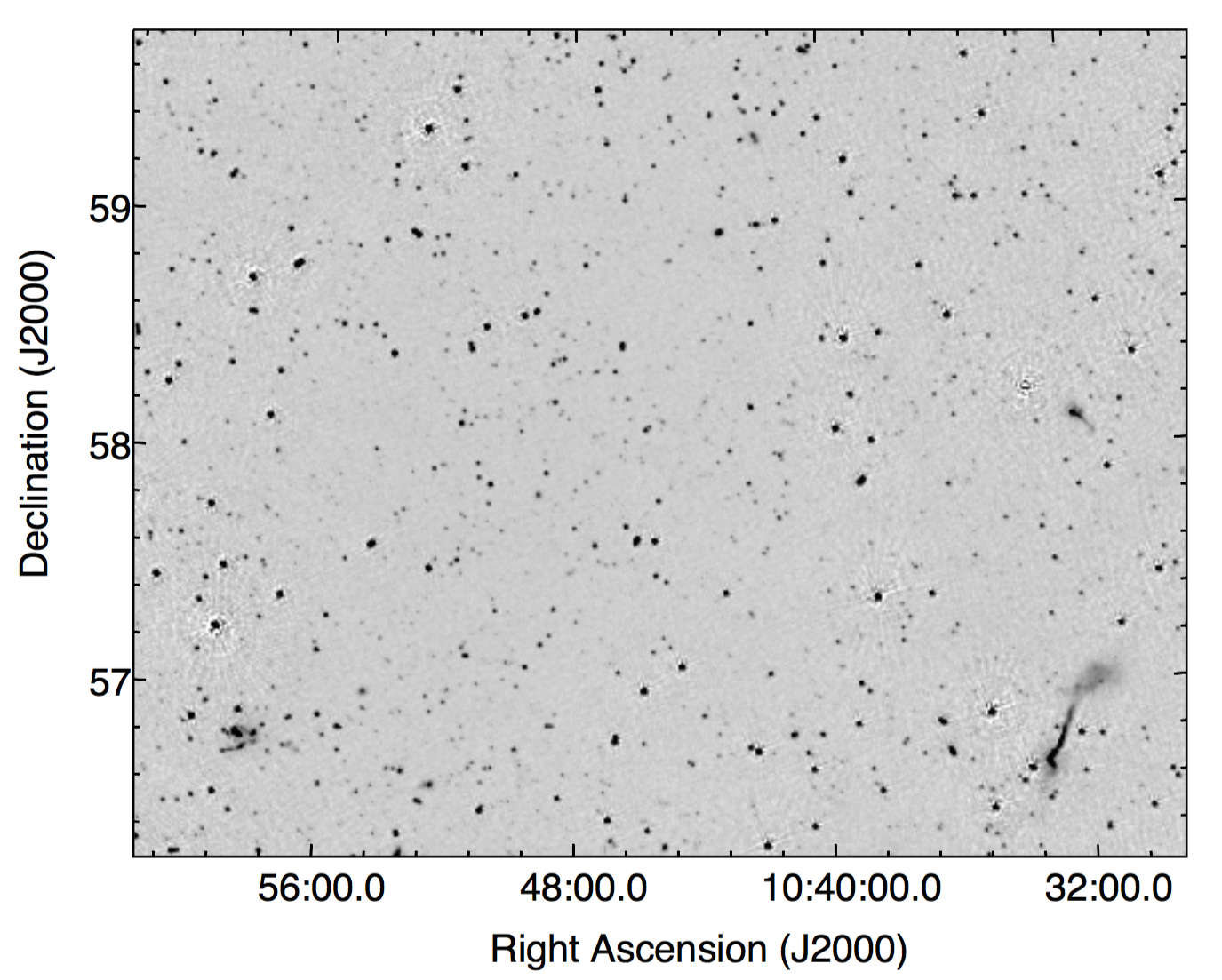}

\caption{LOFAR images of the Lockman Hole at 150 MHz. The top image has a resolution of 18.6$\times$14.7 arcsec and a noise level of 0.15 mJy/beam. The bottom image has a resolution of 45$\times$45 arcsec and a noise level of 1.2 mJy/beam.  }

\label{fig:spectrum}
\end{figure}

\section{Selecting dying radio galaxies}

The identification of dying radio galaxies is not trivial. The classical prototype of a dying radio galaxy has relaxed morphology, without compact components like core, hot-spots or jets. The radio spectrum is expected to be steep ($\rm  \alpha\gtrsim1.2 \ where \ F_\nu=\nu^{-\alpha} $) according to radiative cooling models (Pacholczyk 1970). We should remember though, that active radio galaxies show a variety of characteristics and evolutionary histories and therefore we also expect their remnant phase to cover a wide range of properties based on the radio galaxy conditions before the switch off. For example, they can totally lack nuclear radio emission but they can also show weak radio cores if the central activity has switched off gradually (Saripalli et al. 2012). Adiabatic losses can be negligible or significant depending on whether the lobes are in pressure equilibrium with the surrounding medium at the end of the active phase or not (Hardcastle \& Krause 2013). Also, depending on whether the jets have switched off early or late, the integrated radio spectrum of the source can develop steep spectral indices at low frequency or not (Brienza et al. 2016).

In this context, it is clear that the identification of remnant radio galaxies is challenging. We think therefore, that the use of a single selection criterion can lead to a biased representation of the remnant population. To date, most of the searches have been performed by cross-matching publicly available surveys at different frequencies and have used steep spectral indices as selection criterion. For example, Parma et al. (2007) selected a sample of nearby dying sources by cross matching the 327-MHz Westerbork Northern Sky Survey (WENSS, Rengelink et al. 1997) with the 1.4-GHz NVSS. Dwarakanath \& Kale (2009) and van Weeren et al. (2009) used the same technique combining the 74-MHz VLA Low-Frequency Sky Survey (VLSS, Cohen et al. 2007) and 1400-MHz NVSS. Few authors have also looked for steep spectral index sources in single fields with dedicated high-sensitivity observations (Sirothia et al. 2009; Afonso et al. 2011, van Weeren et al. 2014). By using this criterion, though, we are systematically missing sources that are inactive but whose spectral break frequency still has not reached the low frequencies. This class of objects includes radio galaxies with very low magnetic fields and those with very short active phases (e.g. Brienza et al. 2016). To overcome this simplistic selection criterion, Murgia et al. (2011) suggested to use the spectral curvature ($\rm SPC = \alpha_{high}-\alpha_{low}$) to trace the evolutionary phases of radio sources. Finally, searches based on morphological criteria have also been used by Saripalli et al. (2012) looking for relaxed morphologies and absent compact features. 

\subsection{The search in the Lockman Hole}

In light of what was explained in the previous section we started a search in the Lockman Hole at 150 MHz based on a combination of morphological and spectral selection criteria. In this way we aim at avoiding selection biases towards any specific class of dying sources. In the following we present a summary of the preliminary results obtained using the different criteria:

\begin{itemize}

\item The morphology selection has been performed on the LOFAR highest resolution image via visual inspection using the following criteria: (i) extended size, (ii) relaxed morphology, (iii) low surface brightness, (iv) weak or absent compact components. Another useful criterion to constrain the level of activity of the radio AGN is the core prominence (i.e. ratio between the core power at high frequency and the power of the extended emission at low frequency; Giovannini et al. 1988). However, this requires deep high-frequency and high-resolution data that are currently not available. We used the Faint Images of the Radio Sky at Twenty-cm survey (FIRST, Becker et al. 1994) to check for the presence of any higher resolution counterpart but due to the low sensitivity of the survey (1 mJy/beam detection limit) we cannot derive definitive conclusions from the non-detections. This search resulted in about 10 candidate dying radio galaxies. Few examples of these sources are shown in Fig. 2.

\begin{figure}
 \centering

   {\includegraphics[width=7cm]{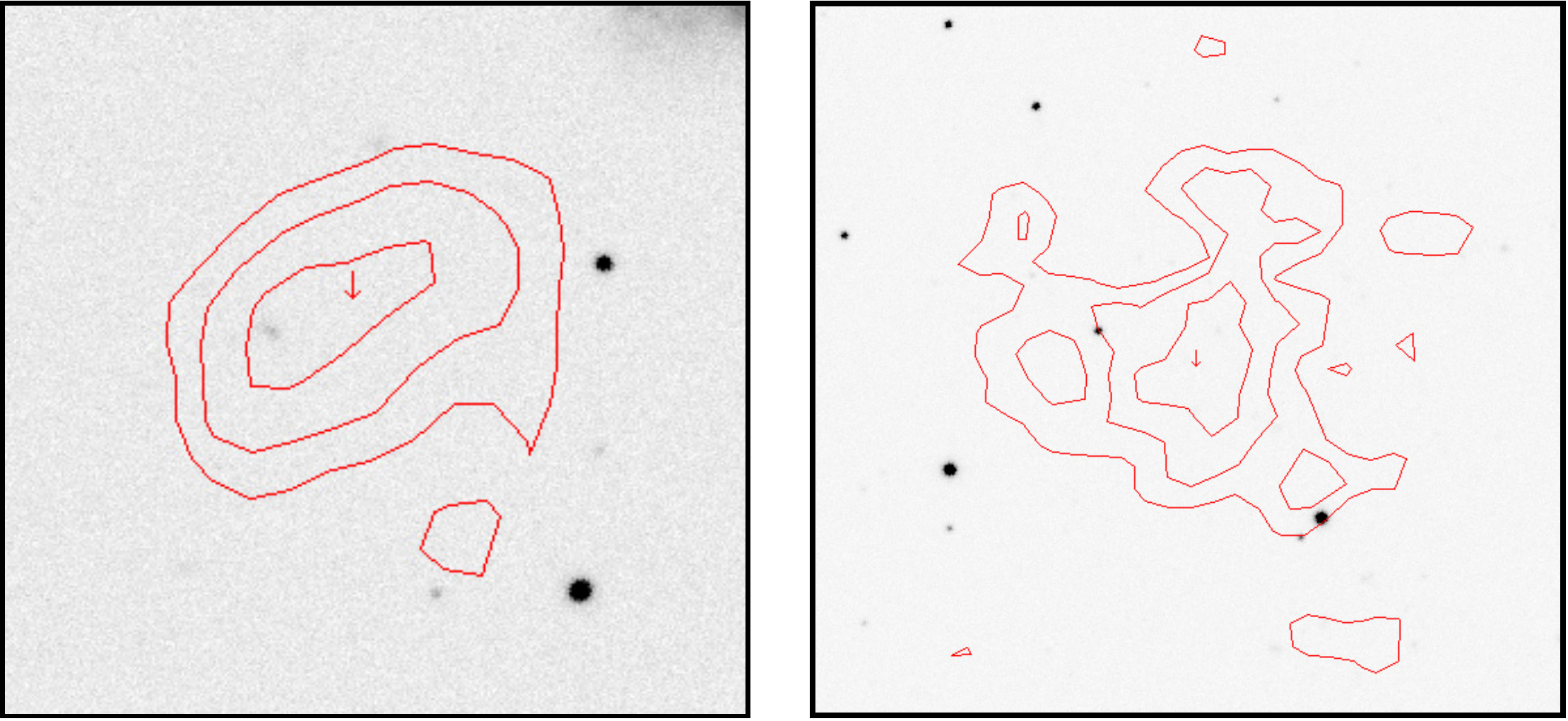}}
\caption{Example of sources with relaxed morphology, low surface brightness and without FIRST counterpart.. LOFAR contours (levels: 3, 4, 5 $\times$ $\sigma$) are overlaid on the SDSS image.  }
 \end{figure}
 
\item For the steep spectral index search we performed two different analysis at different flux limits by using the LOFAR high resolution catalogue and low resolution catalogue combined with the WSRT 1.4 GHz mosaic and with the NVSS respectively (see Sec. 2). The difference in sensitivity between the two 1.4 GHz data is a factor 4.5, meaning that with NVSS we are probing the more powerful radio galaxy population while with the WSRT observation we are mostly sensitive to low-power radio sources. According to spectral ageing models the radio spectrum of active sources is expected to be a broken power law with spectral index $\rm \alpha_{inj}\sim 0.5-0.7$ below a break frequency $\rm \nu_{break}$ and $\rm \alpha= \alpha_{inj}+0.5$ above. Therefore we expect dying sources to show spectral indices $\gtrsim$ 1.2.

In the high resolution catalogue we found that $\sim$7$\% $ of the sources have spectra steeper than 1.2. Moreover, we found that $\sim$ 25$\%$ of the resolved sources (with fitted sizes >26 arcsec) are steep. In the low resolution catalogue we found that $\sim$5.8$\% $ of the sources in the catalogue have spectra steeper than 1.2 and that $\sim$ 35$\%$ of the resolved sources (with fitted sizes >64 arcsec) are steep.  Despite a factor 4.5 difference in sensitivity the two catalogues show very much comparable results (Fig. 3).

\begin{figure}
 \centering

   {\includegraphics[width=16cm]{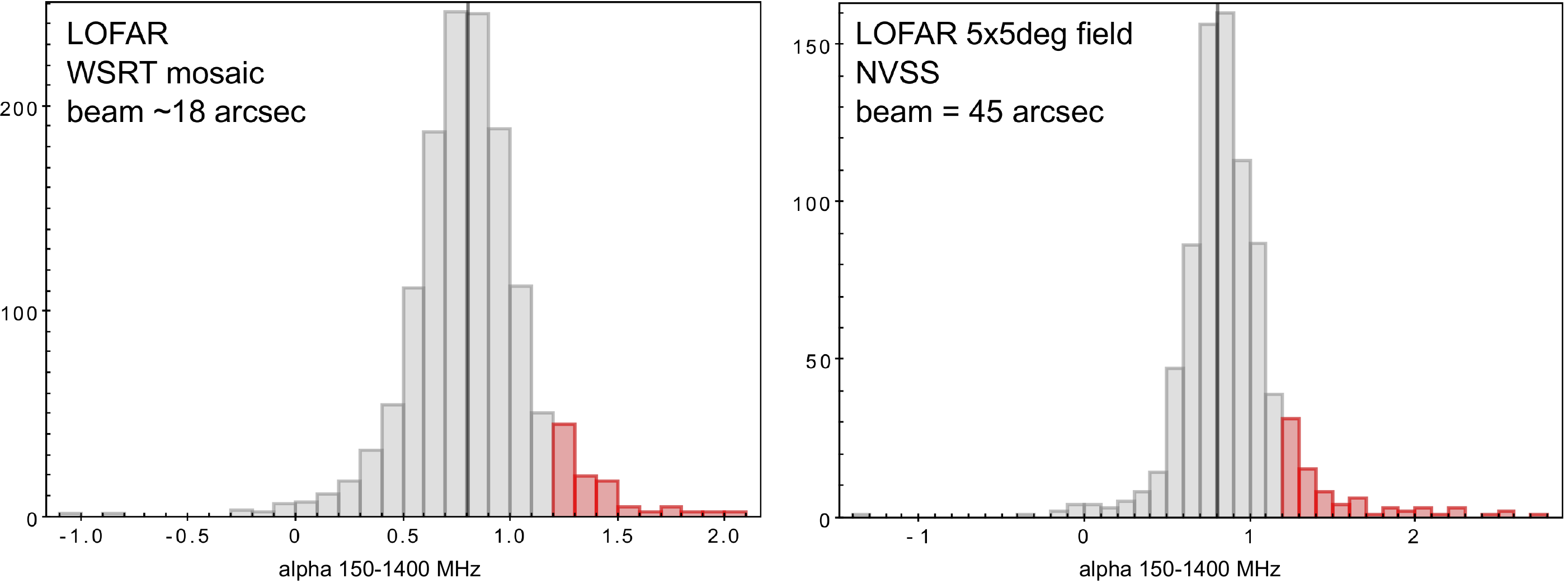}}
\caption{Spectral index distribution between 150 and 1400 MHz where  $\rm F_\nu=\nu^{-\alpha} $. \textit{Left} - High-resolution catalogue obtained by cross-matching the LOFAR catalogue at 18.6$\times$14.7 arcsec resolution with the WSRT 1400-MHz mosaic catalogue (see Sec. 2). \textit{Right} - Low-resolution catalogue obtained by cross-matching the entire LOFAR catalogue at 45 arcsec resolution with the NVSS (see Sec. 2). }
 \end{figure}

\item Following the work done by Murgia et al. (2011) we investigated the possibility of using the spectral curvature criterion by combining the available public radio catalogues. This criterion is aimed at selecting sources whose global spectrum is not so steep to be included in the steep spectral index sample but shows a visible curvature suggesting the central AGN being switched off. 
To perform this, we first cross matched the original low resolution LOFAR catalogue with the WENSS at 327 MHz and then with the NVSS. 
In this way, we obtained a catalogue of 790 sources. We computed the spectral curvature as SPC = $\rm \alpha_{high}-\alpha_{low}$ using $\rm \alpha_{low}=\alpha^{150}_{327}$ and $\rm \alpha_{high} = \alpha^{327}_{1400}$.
We classified as candidates those sources having SPC>0.5. With this method we were able to select only 4 candidates. The reason for the paucity of sources extracted is likely connected with both the low sensitivity of the higher frequency surveys and the lack of data at frequencies higher than 1.4 GHz where the break frequency is most likely to occur.

\end{itemize}

\section{Conclusions and future plans}

In these proceedings we have presented a summary of the search of dying radio galaxies that we are performing in the LOFAR image of the Lockman Hole at 150 MHz. We have explained why we think it is important to use complementary approaches to probe different classes of dying sources. Our preliminary results, show that the total fraction of candidate dying radio sources is <~6~-~8~\% of the entire radio source population and is dominated by steep spectrum sources. The current main limitations in the search with other methods are connected to the absence of high frequency deep radio data. For this reason high-sensitivity surveys at 1.4 GHz like APERTIF (Oosterloo et al. 2009) will be crucial for these kind of studies. Confirmation of the nature of the selected candidate sample will be presented in a future work. In order to deeper understand the nature of our observational results we are also performing statistical simulations of source populations based on radio galaxy evolution models developed by Godfrey et al. (in prep.). The comparison of the observed catalogues with simulated mock catalogues will help us constrain the importance of different mechanisms on the dying radio galaxy luminosity evolution. The observational and theoretical framework developed in this work will be later expanded to other fields of the LOFAR Tier1 survey and finally to the whole northern sky. Automated algorithms like the BRATS survey module (Harwood et al., this volume) will be a key to handle these very extended searches.

\section*{Acknowledgements}

We thank the organizers of the conference "The many facets of extragalactic radio surveys: towards new scientific challenges" and RadioNet for the financial support. The research leading to these results has received funding from the European Research Council under the European Union's Seventh Framework Programme (FP/2007-2013) / ERC Advanced Grant RADIOLIFE-320745. LOFAR, the Low Frequency Array designed and constructed by ASTRON (Netherlands Institute for Radio Astronomy), has facilities in several countries, that are owned by various parties (each with their own funding sources), and that are collectively operated by the International LOFAR Telescope (ILT) foundation under a joint scientific policy.

\end{document}